\documentclass[preprint,aps,showpacs,showkeys,tightenlines,superscriptaddress]{revtex4}
\usepackage{epsfig}
\usepackage{amsmath}
\newcommand{\vi}[1]{\mbox{\boldmath $#1$}}
\newcommand{\vis}[1]{\mbox{\boldmath ${\scriptstyle #1}$}}
\begin{document}
\title{Systematic analysis of reaction cross sections 
of carbon isotopes}
\author{W. Horiuchi}
\affiliation{Graduate School of Science and Technology, Niigata
University, Niigata 950-2181, Japan}
\author{Y. Suzuki}%
\affiliation{Department of Physics and Graduate School of Science and 
Technology, Niigata University, Niigata 950-2181, Japan}
\author{B. Abu-Ibrahim}
\affiliation{RIKEN Nishina Center, Wako-shi, Saitama 351-0198, Japan}
\affiliation{Department of Physics, Cairo University, 
Giza 12613,Egypt}
\author{A. Kohama}
\affiliation{RIKEN Nishina Center, Wako-shi, Saitama 351-0198, Japan}

\date{\today}

\begin{abstract}
We systematically analyze total reaction cross sections 
of carbon isotopes with $N=6$--$16$ on a $^{12}$C target 
for wide range of incident energy. 
The intrinsic structure of the carbon isotope is described 
by a Slater determinant generated from a phenomenological 
mean-field potential, 
which reasonably well reproduces the ground state properties 
for most of the even $N$ isotopes. 
We need separate studies not only for odd nuclei but also 
for $^{16}$C and $^{22}$C. The density of the 
carbon isotope is constructed by eliminating the effect of 
the center of mass motion. 
For the calculations of the cross sections, 
we take two schemes: one is the Glauber approximation, and 
the other is the eikonal model using a global optical potential. 
We find that both of the schemes successfully reproduce  
low and high incident energy data on 
the cross sections of $^{12}$C, $^{13}$C and $^{16}$C on $^{12}$C.  
The calculated reaction cross sections 
of $^{15}$C are found to be considerably smaller than  
the empirical values observed at low energy. 
We find a consistent parameterization 
of the nucleon-nucleon scattering amplitude, differently 
from previous ones. 
Finally, we predict the total reaction cross section of $^{22}$C
on $^{12}$C. 
\end{abstract}

\maketitle

\section{Introduction}

The structure of carbon isotopes has recently attracted much 
attention as experimental information accumulated 
toward its neutron dripline. 
The topics discussed include, for example,  
the subshell closure of $N$=$14$ and $N$=$16$ 
and the anomalously small 
$E2$ transition strength observed in $^{16}$C~\cite{imai,elekes}. 
These issues are 
closely related to the competition of $0d_{5/2}$ 
and $1s_{1/2}$ neutron orbits. In fact, they play a 
predominant role in determining the ground state structure of 
the carbon isotope with $N>8$. 

The momentum 
distribution of a $^{15}$C fragment in the breakup of $^{16}$C 
suggests that the last neutrons in $^{16}$C occupy both the 
$0d_{5/2}$ and $1s_{1/2}$ orbits~\cite{yamaguchi}, which is 
consistent with recent $^{14}$C+$n$+$n$ three-body 
calculations~\cite{suzukic16,horiuchi}. 
The $1s_{1/2}$ orbit plays a vital 
role in forming a one-neutron halo structure of 
$^{19}$C~\cite{nakamura,maddalena}. 
If the subshell closure of $N=14$ is a good approximation in the 
carbon isotopes, the $0d_{5/2}$ orbits are fully occupied 
in the nucleus $^{20}$C. Adding one more neutron to $^{20}$C 
leads to no particle-bound system, but $^{22}$C, getting 
one more neutron, becomes bound. $^{22}$C is thus a Borromean 
nucleus. The structure of $^{22}$C has recently been studied 
by two (W.H. and Y.S.) of the present authors 
in the three-body model of $^{20}$C+$n$+$n$~\cite{c22}. 

A molecular picture in $^{14}$C is investigated in the framework of 
three $\alpha$-particles plus two neutrons~\cite{itagaki}. 
The deformation of the carbon
isotopes is also discussed to have a strong 
$N$-dependence~\cite{kanada,sagawa}. 
The properties of the carbon isotopes are reviewed in 
Ref.~\cite{brown} based on the mean field and shell-model 
configuration mixing models. 
 
How do such nuclear structures affect reaction data? 
Nowadays, the data on total reaction or interaction 
cross sections have accumulated 
particularly for light nuclei. In the case of the carbon isotopes, 
for example, the interaction cross section has been measured up to 
$^{20}$C around 700$\sim$960 $A\,$MeV incident energy~\cite{ozawa}. 
Since these cross sections reflect the size of nuclei, it is 
interesting to analyze the cross sections in a systematic manner. 

The purpose of the present study is a systematic analysis of 
the total reaction cross sections of carbon isotopes on a 
$^{12}$C target using two reaction models which enable us to 
go beyond a folding model; the Glauber model~\cite{glauber} and 
the eikonal approximation~\cite{yos,slyv} with the use of 
nucleon-$^{12}$C optical potentials. 
This study is also motivated by an ongoing measurement of 
the reaction cross section of $^{22}$C~\cite{tanaka}. 
Such a measurement looks quite challenging because the 
production rate of $^{22}$C is expected to be small. We will perform 
a simple, consistent, {\it ad hoc} parameter-free analysis. 
The systematics will offer an interesting interplay 
between nuclear structure models and the reaction models. 

The input parameters on nucleon-nucleon scatterings 
needed in the Glauber calculation is carefully 
assessed using available 
$^{12}$C+$^{12}$C reaction cross section data. The wave function 
of a carbon isotope is first generated from a Slater determinant 
whose nucleon orbits are built from phenomenological 
mean-field potentials, and the corresponding 
neutron and proton densities, 
with its centre of mass (c.m.) motion being 
taken into account properly, 
are constructed for the calculation of  
the total reaction cross section. 
The asymptotic form of the wave function 
is carefully described by the use of empirical nucleon 
separation energies 
as it is important for the cross section calculation, particularly 
for a spatially extended system. 
A comparison with  experimental cross sections will 
immediately reveal a successful or unsuccessful case. 
To resolve the discrepancy between theory and experiment, 
one has to go beyond the simple mean-field model and two types 
of dynamical models are performed to obtain an improved density. 
One is a core+$n$ model for an odd $N$ nucleus, and the other 
is a core+$n$+$n$ model for $^{16}$C and $^{22}$C.  

The organization of the present paper is as follows: The 
reaction models for a calculation of reaction cross sections 
between nuclei are presented in the next section. A simple 
formula is given in Sect.~\ref{glauber} in the framework of 
the Glauber theory, and the other method using an optical potential 
is explained in Sect.~\ref{NTsec}. The reaction cross section 
of $^{12}$C+$^{12}$C is tested by these formulas in a wide range of 
incident energy. In Sect.~\ref{dens.slater} the phenomenological 
mean-field potential is prescribed for generating 
the Slater determinant, and the c.m. motion is removed 
in order to obtain the intrinsic density 
which is used in the reaction 
calculation. The mean radius of the matter distribution is compared 
to the empirical value.  
The nuclear structure model is extended to the dynamical model in 
Sect.~\ref{dens.dynamic}. A core+$n$ model is applied to the odd $N$ 
isotopes in Sect.~\ref{dens.coren}, where the difference in 
the densities between the dynamical model and the Slater determinant 
is discussed. The binding energy and the matter 
size of $^{22}$C are studied in the three-body model of 
$^{20}$C+$n$+$n$ in Sect.~\ref{carbon22} and the densities of 
the core+$n$+$n$ model are presented in Sect.~\ref{dens.corenn}. 
Section~\ref{result} presents the results of reaction cross section 
calculations; the cases of $^{12-20}$C in Sect.~\ref{rcs.c12-20} and 
the $^{22}$C+$^{12}$C reaction in Sect.~\ref{rcs.c22}. Summary is 
drawn in Sect.~\ref{summary}. A method of calculation of 
two-particle distribution functions is given in Appendix.

\section{Model for a reaction cross section calculation}

In this section, we describe our reaction models 
for analyzing reaction cross sections between nuclei.  
A simple formula is given in Sect.~\ref{glauber} 
in the framework of 
the Glauber theory, and the other method using an optical potential 
is explained in Sect.~\ref{NTsec}. 
These methods are complimentary 
to each other for a $^{12}$C target, 
but only the former can be applied for a proton target. 
With these calculations in two ways, 
we can find a reliable parameterization 
of the $NN$ interaction for a wide energy range, 
which is important to proceed to
the case of a proton target in our future work. 

\subsection{GLAUBER FORMALISM}
\label{glauber}

The reaction cross section for a projectile-target collision 
is calculated by integrating the 
reaction probability with respect to the impact parameter {\vi b};
\begin{equation}
   \sigma_{{\rm R}}=\int{d{\vi b}\,
   \left(1-\big|{\rm e}^{i\chi({\vis b})}\big|^{2}\right)},
\label{reac.form}
\end{equation}
where the phase shift function $\chi$ is expressed, in the Glauber 
model~\cite{glauber}, through the $NN$ profile function 
$\Gamma_{NN}$ by
\begin{equation}
   {\rm e}^{i\chi({\vis b})}=\langle\Psi_{0}\Theta_{0}
   \vert\prod_{i\in \rm P}\prod_{j\in \rm T}
   \left\{1-\Gamma_{NN}({\vi s}_i-{\vi t}_j+{\vi b}) \right\}\vert
   \Psi_{0}\Theta_{0}\rangle.
\label{psfunc}
\end{equation}
Here $\Psi_{0}$ ($\Theta_{0}$) is the intrinsic wave function of 
the projectile (target) with its c.m. part being removed, 
${\vi s}_i$ is the two-dimensional vector 
of the projectile's single-particle coordinate, ${\vi r}_i$, 
measured from the projectile's c.m. 
coordinate, and ${\vi t}_i$ is defined for the target nucleus in 
a similar way. The profile function $\Gamma_{NN}$ is 
usually parameterized in the form; 
\begin{equation}
   \Gamma_{NN}({\vi b})=\frac{1-i\alpha}{4\pi \beta}\,
   \sigma_{NN}^{\rm tot}\,
   {\rm exp}\Big(-\frac{{\vi b}^{2}}{2\beta}\Big), 
\label{profilefn}
\end{equation}
where $\sigma_{NN}^{\rm tot}$ is the total cross section for 
${\it NN}$ collisions, $\alpha$ 
the ratio of the real to the imaginary part of the 
${\it NN}$ scattering amplitude, 
and $\beta$ the slope parameter of the ${\it NN}$ elastic 
differential cross section.  

As seen in Eq.~(\ref{psfunc}), the 
calculation of the phase shift function requires a multi-dimensional 
integration. Though the integration can be performed using the 
Monte Carlo 
technique even for sophisticated 
wave functions~\cite{varga}, it is fairly involved in general, so 
it is often approximately evaluated in the optical limit
approximation (OLA) using the intrinsic densities 
of the projectile (target) nuclei, 
$\rho_{\rm P}$ ($\rho_{\rm T}$), as follows: 
\begin{equation}
   {\rm e}^{i\chi_{\rm OLA}({\vis b})}
    ={\rm exp}\left\{-\int\!\!{\int{d{\vi r} d{\vi r^{\prime}} 
   \rho_{\rm P}({\vi r})\rho_{\rm T}({\vi r^{\prime}})
   \Gamma_{NN}({\vi s}-{\vi t}+{\vi b})}}\right\}.
\label{ola}
\end{equation}

Another approximation is proposed in Ref.~\cite{utility} 
by two (B.A-I. and Y.S.) of the present authors to 
calculate the reaction cross sections 
using the same input as in the OLA. 
The essence of the approximation is 
to consider, as an intermediate step, a phase shift function for 
the nucleon-nucleus scattering. 
With the introduction of the profile function $\Gamma_{NT}$ 
for the nucleon-target (NT) scattering, 
the phase shift function of OLA, Eq.~(\ref{ola}), is 
replaced by $\bar{\chi}$ as
\begin{eqnarray}
   {\rm e}^{i\bar{\chi}({\vis b})}&=&
   \langle \Psi_{0}
   \vert \prod_{i\in \rm P}
   \left\{1-\Gamma_{NT}({\vi s}_i+{\vi b}) \right\}\vert
   \Psi_{0}\rangle
\nonumber \\
&\approx& {\rm exp}\left(-\int{ }d{\vi r} 
   \rho_{\rm P}({\vi r})\Gamma_{NT}({\vi s}+{\vi b})
   \right).
\label{gntola}
\end{eqnarray}
We here adopt two methods to calculate the $\Gamma_{NT}$: One is to 
calculate the $\Gamma_{NT}$ using an appropriate optical potential 
as shown in the next subsection. The other is to use the Glauber 
theory as  
\begin{eqnarray}
   \Gamma_{NT}({\vi b})
   &=&1- \langle \Theta_{0}
   \vert \prod_{j\in \rm T}
   \left\{1-\Gamma_{NN}({\vi b}-{\vi t}_j) \right\}\vert
   \Theta_{0}\rangle
\nonumber \\
   &\approx&1-{\rm exp}\left(-\int{ }d{\vi r}' 
   \rho_{\rm T}({\vi r}')\Gamma_{NN}({\vi b}-{\vi t})
   \right).
\end{eqnarray}
Substituting this expression into Eq.~(\ref{gntola}) leads us to 
\begin{equation}
   {\rm e}^{i\bar{\chi}({\vis b})}=
   {\rm exp}\left[-\int{ }d{\vi r} 
   \rho_{\rm P}({\vi r})\left\{
   1-{\rm exp}\left(-\int{ }d{\vi r}' 
   \rho_{\rm T}({\vi r}')\Gamma_{NN}({\vi s}-{\vi t}+{\vi b})
   \right)\right\}\right] .
\label{gpteff}
\end{equation}
This formula is found to give better results 
than those of the OLA~\cite{utility,abu00}. 
Though only the leading term in the cumulant expansion is taken 
into account to derive Eq.~(\ref{gpteff}), 
this approximation includes higher order 
corrections which Eq.~(\ref{ola}) does not contain~\cite{abu00}. 
Since the role of the projectile and the target is 
interchangeable in the calculation of 
the reaction cross section, it may be possible to 
symmetrize Eq.~(\ref{gpteff}) as follows:
\begin{eqnarray}
   {\rm e}^{i\bar{\chi}({\vis b})}&=&
   {\rm exp}\left[-\frac{1}{2}\int{ }d{\vi r} 
   \rho_{\rm P}({\vi r})\left\{
   1-{\rm exp}\left(-\int{ }d{\vi r}' 
   \rho_{\rm T}({\vi r}')\Gamma_{NN}({\vi s}-{\vi t}+{\vi b})
   \right)\right\}\right] \nonumber \\
   &\times& 
   {\rm exp}\left[-\frac{1}{2}\int{ }d{\vi r}' 
   \rho_{\rm T}({\vi r}')\left\{
   1-{\rm exp}\left(-\int{ }d{\vi r} 
   \rho_{\rm P}({\vi r})\Gamma_{NN}({\vi t}-{\vi s}+{\vi b})
   \right)\right\}\right] .
\label{gptsym}
\end{eqnarray}  
This approximation is called NTG hereafter, which stands for the 
NT profile function in the Glauber model. 

% Fig.1
\begin{figure}[b]
\epsfig{file=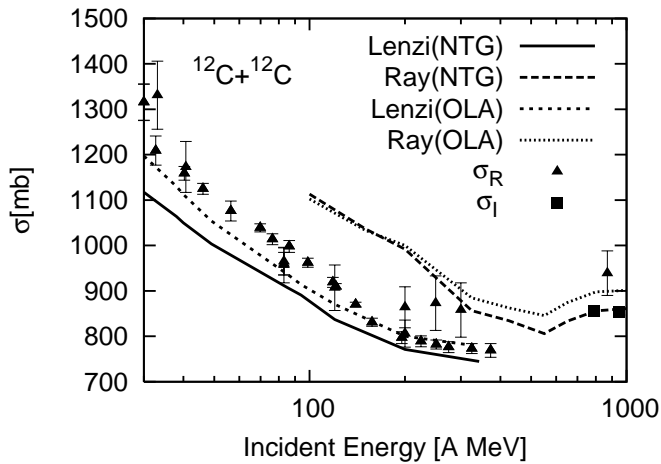,width=9.01cm,height=6.307cm}
\caption{Reaction cross sections of $^{12}$C 
on a $^{12}$C target calculated with the parameters 
of Refs.~\cite{lenzi,ray}.  The experimental data are taken from 
Refs.~\cite{takechi,perrin,zhang,fang00,kox,zheng,hostachy,jaros} 
for the reaction cross section $\sigma_{\rm R}$ and from 
Ref.~\cite{ozawa} for the 
interaction cross section $\sigma_{\rm I}$. }
\label{ccreac}
\end{figure}

The parameters of $\Gamma_{NN}$ are taken from 
Refs.~\cite{lenzi,ray}. In the latter case~\cite{ray}
the parameters are given for 
the $pp$ and $pn$ collisions separately, 
but here we use the mean values because the target 
nucleus is $^{12}$C whose proton and neutron densities are 
virtually the same to good accuracy. 
In Fig.~\ref{ccreac} we compare the numerical results 
obtained using these parameters
with the experimental data of $^{12}$C+$^{12}$C 
total reaction cross sections. 
Here the intrinsic density of $^{12}$C is obtained from the 
procedure which will be explained in the next section.  
It is found that both the parameters give quite different cross 
sections at 100$\sim$300 $A\,$MeV. Apparently the cross 
sections obtained with the parameters of Ref.~\cite{ray} are 
too large, while those with the parameters of Ref.~\cite{lenzi} 
tend to be a little smaller than experiment. 

The $NN$ profile function could be 
subject to change from that of the free space 
especially at lower energies because of the effects due to the 
Pauli blocking and the Fermi motion of the 
nucleons~\cite{giacomo}. Warner {\it et al.} 
studied the in-medium effect on the reaction cross section 
by modifying the free $\sigma_{NN}^{\rm tot}$~\cite{warner}. 
Takechi {\it et al.} have recently reported that 
taking into account the Fermi motion leads to a 
significant change in the $\sigma_{NN}^{\rm tot}$
values, which is vital to reproduce the reaction cross sections 
at lower energies~\cite{fukuda}.  

Here we take a simpler route: First, 
we note that the total elastic cross section $\sigma_{NN}^{\rm el}$ 
of the $NN$ collision is given by 
\begin{equation}
   \sigma_{NN}^{\rm el}=\frac{1+\alpha^2}{16\pi \beta}
   \left(\sigma_{NN}^{\rm tot}\right)^2
\label{el-total}
\end{equation}
for the profile function of Eq.~(\ref{profilefn})~\cite{ogawa}.
Then, for $E < 300\, A\,$MeV where only 
the elastic scattering is energetically possible as the pion
production threshold is closed, we expect that 
the relation of 
$\sigma_{NN}^{\rm el}=\sigma_{NN}^{\rm tot}$ should hold 
from the unitarity of the $NN$ collision. 
Employing the parameters of Ref.~\cite{ray} yields 
$\sigma_{NN}^{\rm el}=17, \, 7$ and $3\,$mb at $E=100, \, 150$ 
and $200\,$MeV, respectively, which are far smaller than the  
$\sigma_{NN}^{\rm tot}$ values at the corresponding energies.
We, instead, choose the $\beta$ value for $E < 300\, A\,$MeV as 
\begin{equation}
   \beta=
   \frac{1+\alpha^2}{16\pi }\sigma_{NN}^{\rm tot}
\end{equation}
to satisfy the equality 
of $\sigma_{NN}^{\rm el}=\sigma_{NN}^{\rm tot}$. 
For $E > 300\, A\,$MeV where the equality breaks down, 
the $\beta$ values are determined from Eq.~(\ref{el-total}) 
using the experimental 
values of $\sigma_{NN}^{\rm el}=\frac{1}{2}(\sigma_{pp}^{\rm el}+
\sigma_{pn}^{\rm el})$~\cite{pdg}. 
Some of the $\alpha$ parameters of Ref.~\cite{lenzi} are also 
modified to follow the systematics of Ref.~\cite{ray}.

\begin{figure}[h]
\epsfig{file=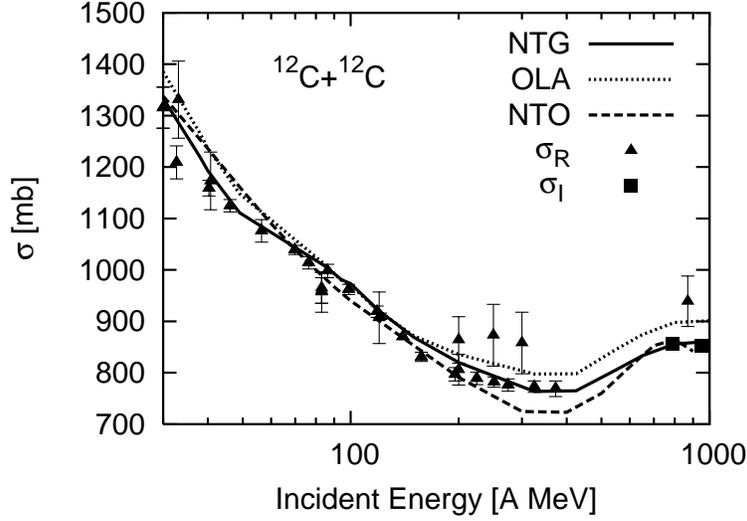,width=10.2cm,height=7.14cm}
\caption{Comparison of the reaction cross sections of $^{12}$C 
on a $^{12}$C target between theory and experiment. The input 
parameters for $\Gamma_{NN}$ are taken from Table~\ref{nnprofile} 
in the NTG and OLA calculations, while those for $\Gamma_{NT}$ 
are taken from the global optical potential of Ref.~\cite{cooper} 
in the NTO calculation. At energies less than 100 $A\,$MeV the
correction due to the deviation from the straight-line
trajectory, though negligibly small, 
is included in the NTO calculation. 
See the caption of Fig.~\ref{ccreac} for the experimental data.}
\label{c12-c12reac}
\end{figure}

Table~\ref{nnprofile} lists the parameters of the $NN$ profile 
function used in the present study. The $^{12}$C+$^{12}$C 
reaction cross sections 
calculated using these parameters are displayed 
by solid (NTG) and dotted (OLA) lines in Fig.~\ref{c12-c12reac}. 
We find that the modified parameter set reproduces very well the 
experiment in the whole energy region. 
The NTG phase shift function is found to reproduce the 
cross section better than the OLA. 
We thus conclude that both the calculated density of 
$^{12}$C and the parameter set of $\Gamma_{NN}$ are qualified  
for a systematic analysis of the reaction 
cross section of the carbon isotopes on a $^{12}$C target.

\begin{table}[t]
\caption{Parameters of the $NN$ profile function. $E$ is the 
projectile's incident energy. Some parameters are modified from 
the original values of Refs.~\cite{lenzi,ray}. See the text for 
detail. }
\label{nnprofile}
\begin{center}
\begin{tabular}{cccc}
\hline\hline
$E$ & $\sigma_{NN}^{\rm tot}$ & $\alpha$ & $\beta$ \\
($A\,$MeV) & (fm$^2$) &  &  (fm$^2$)  \\
\hline
30 &  19.6 &  0.87 &  0.685 \\
38 &  14.6 &  0.89 &  0.521 \\
40 &  13.5 &  0.9  &  0.486 \\
49 &  10.4 &  0.94 &  0.390 \\
85 &  6.1  &  1.37 &  0.349 \\
94 &  5.5  &  1.409&  0.327 \\
100&  5.295&  1.435&  0.322 \\
120&  4.5  &  1.359&  0.255 \\
150&  3.845&  1.245&  0.195 \\
200&  3.45 &  0.953 &  0.131 \\
325&  3.03 &  0.305&  0.075 \\
425&  3.03 &  0.36 &  0.078 \\
550&  3.62 &  0.04 &  0.125 \\
650&  4.0  & $-$0.095&  0.16  \\
800&  4.26 & $-$0.07 &  0.21  \\
1000& 4.32 & $-$0.275&  0.21 \\
\hline\hline
\end{tabular}
\end{center}
\end{table}

%Fig.2

\subsection{NUCLEON-NUCLEUS DATA AS A BASIC INPUT}
\label{NTsec}

In this subsection, we briefly present a method developed 
in Ref.~\cite{utility} 
for describing nucleus-nucleus scattering 
using an optical potential for the nucleon-nucleus 
elastic scattering.

Denoting the NT optical potential by $V_{NT}$, we define 
the corresponding phase shift function $\chi_{NT}$ as 
\begin{equation}
   \chi_{NT}({\vi b})=-\frac{1}{\hbar v}
   \int_{-\infty}^{\infty} dz\,V_{NT}({\vi b}+ z\hat{\vi z}),
\end{equation}
where $v$ is the incident velocity of the NT relative motion. 
Now we express the NT profile function as 
\begin{equation}
   \Gamma_{NT}({\vi b})=1-{\rm e}^{i\chi_{NT}({\vis b})}.
\label{eik}
\end{equation}
The substitution of Eq.~(\ref{eik}) into Eq.~(\ref{gntola}) yields 
another formula to calculate the optical phase shift function. 
We call this approximation NTO, which stands for the NT formalism
with the optical potential. Similarly to NTG, the reaction 
cross section given by NTO includes higher order terms which 
are missing in the reaction cross section calculated with a 
folding model.  In the latter model the phase shift function 
$\chi_f$ is simply given by 
\begin{equation}
   {\rm e}^{i\chi_{f}({\vis b})}
   ={\rm exp}\left( i\int \, d{\vi r} \, \rho_{\rm P}({\vi r})
   \chi_{NT}({\vi b}+{\vi s})\right).
\label{foldig.psf}
\end{equation}
 
The needed input for NTO is the projectile's 
intrinsic density and the optical potential $V_{NT}$ 
at a given energy. 
As $V_{NT}$ we use the central part of the 
global optical potential EDAD-fit3 (GOP)~\cite{cooper}, which  
is determined by a Dirac phenomenology. 
This potential, together with the other EDAD sets, gives 
a good fit to $p$+$^{12}$C elastic scattering and  
reaction cross section data in the incident energy of 30 MeV
to 1 GeV. It should be noted, however, that 
the EDAD-fit3 potential predicts slightly smaller reaction 
cross sections than experiment 
in the intermediate energy range of 300--400 MeV. 
This would be due to the lack of data of 
$p$+$^{12}$C elastic scattering differential cross section 
for this energy range. 
We ignore the difference between $p$T and $n$T interactions 
in this study. 

The NTO calculation for $^{12}$C+$^{12}$C reaction 
cross section is shown by dashed line in Fig.~\ref{c12-c12reac}. 
As we see, the agreement between experiment and theory is good. 
The underestimation of the cross section around 300--400 $A\,$MeV 
is probably due to the smaller absorption of the EDAD-fit3 
potential as noted above. 
An advantage of these calculations is that they are parameter free. 
For the energy less than 300 $A\,$MeV, 
the numerical results with NTO as well as with NTG 
agree with the data quite well. 

At energies less than 100 $A\,$MeV, the
correction due to the deviation from the straight-line
trajectory was studied for $^{12}$C+$^{12}$C case.
We used the distance of the 
closest approach in Rutherford orbit in place of 
the asymptotic impact parameter~\cite{charagi}.
This correction is found to be small. 
It decreases the reaction cross section 
by only few percentages at 30 $A\,$MeV. 

For high energy side, 
we note that the reaction cross section calculated 
using NTO slightly decreases at 900 $A\,$MeV. 
This is due to the fact that 
the imaginary part of the GOP reaches its deepest value 
at 800 $A\,$MeV and then decreases by a small amount as 
the energy increases. 

Our results underestimate the data 
of the total reaction cross sections 
at 870 $A\,$MeV~\cite{jaros}. 
The numerical results unexpectedly agree with 
the data of the interaction cross sections 
at 790 and 950 $A\,$MeV~\cite{ozawa}, 
but not with the total reaction cross section. 
Since our result is quite close to 
the one ($\sigma_{\rm R}= 865 \pm 1$ mb) 
obtained from a more sophisticated 
calculation ~\cite{varga}, 
the approximation which we used must be in
appropriate direction. 
Possible uncertainties comes from the parameters 
of NN scattering amplitude and/or the data 
of $\sigma_{\rm R}$ itself. 
In order to clarify the situation, 
a more accurate measurement of such quantities 
at high energy will be useful.

If we believe the data of $\sigma_{\rm R}$ at 870 $A\,$MeV, 
we need a steep increase of the cross section 
from 400 $A\,$MeV 
toward higher energies in order to reproduce it 
while the energy dependence of our results is rather weak. 
As one can see from Fig.1, compared with OLA, 
the NTG, which resums higher order corrections 
coming from the first cumulant as in Eq.~(\ref{gptsym}), 
reduces the magnitude of the cross section for 
the region of the energy higher than 200 $A\,$MeV, 
and causes a weak energy dependence for this energy region.

In contrast to our results, 
a rather strong energy dependence is obtained 
by Iida {\it et al.} based on the black-sphere picture 
of nuclei~\cite{IKO}. 
These authors reproduce the total reaction cross section 
at 870 $A\,$MeV~\cite{jaros} as well as the data 
between 100 and 400 $A\,$MeV 
due to the steep increase of the cross section. 
However, they failed to reproduce the energy dependence of 
low energy side, 
because their picture breaks down for low energy, less than 
around 100 $A\,$MeV. 

Other works, for example, Refs.~\cite{kox,warner}, 
deal with $^{12}$C+$^{12}$C reactions 
of wide range of incident energy, 
and reproduce the reaction cross section at 870 $A\,$MeV.
However, in the energy of 100--400 $A\,$MeV, 
their results agree with the old data~\cite{kox}, 
the larger ones, not the recent smaller ones~\cite{takechi}. 
Therefore, these theoretical results 
overestimate the cross sections in this energy region, 
and the weak energy dependence of their results leads 
to reproducing the reaction cross section at 870 $A\,$MeV.

\section{Density with a Slater determinant}
\label{dens.slater}

Now we discuss the densities of the carbon isotopes, which 
will be applied to the calculation of total 
reaction cross sections.

\begin{table}[t]
\caption{Neutron configurations for the ground states of 
the carbon isotopes. $J^{\pi}$ is the spin-parity 
of the ground state. }
\label{config}
\begin{center}
\begin{tabular}{ccl}
\hline\hline
Nucleus& $J^{\pi}$& configurations \\
\hline
$^{12}$C &$0^+$&$(0s_{1/2})^2(0p_{3/2})^4$ \\
$^{13}$C &$\frac{1}{2}^-$&$(0s_{1/2})^2(0p_{3/2})^4(0p_{1/2})^1$\\
$^{14}$C &$0^+$&$(0s_{1/2})^2(0p_{3/2})^4(0p_{1/2})^2$  \\
$^{15}$C &$\frac{1}{2}^+$
  &$(0s_{1/2})^2(0p_{3/2})^4(0p_{1/2})^2(1s_{1/2})^1$ \\
$^{16}$C &$0^+$&$(0s_{1/2})^2(0p_{3/2})^4(0p_{1/2})^2(0d_{5/2})^2$ \\
$^{17}$C &$-$
  &$(0s_{1/2})^2(0p_{3/2})^4(0p_{1/2})^2(0d_{5/2})^2(1s_{1/2})^1$ \\
  &$-$ &$(0s_{1/2})^2(0p_{3/2})^4(0p_{1/2})^2(0d_{5/2})^3$ \\
$^{18}$C &$0^+$&$(0s_{1/2})^2(0p_{3/2})^4(0p_{1/2})^2(0d_{5/2})^4$\\
$^{19}$C &$(\frac{1}{2}^+)$
  &$(0s_{1/2})^2(0p_{3/2})^4(0p_{1/2})^2(0d_{5/2})^4(1s_{1/2})^1$\\
$^{20}$C &$0^+$
  &$(0s_{1/2})^2(0p_{3/2})^4(0p_{1/2})^2(0d_{5/2})^6$ \\
\hline
\end{tabular}
\end{center}
\end{table}

The intrinsic densities of the carbon isotopes are calculated 
from a phenomenological mean-field potential. 
We assume a Slater determinant for the ground state 
wave function of the carbon isotope. 
Table~\ref{config} lists the neutron configurations assumed for the 
ground states of the carbon isotopes. Some remarks on the 
configurations are made below. Though we assume that the 
last two neutrons occupy the $0d_{5/2}$ orbit for $^{16}$C, 
its ground state is known to contain 
$(1s_{1/2})^2$ and $(0d_{5/2})^2$ configurations nearly 
equally~\cite{yamaguchi,suzukic16}. We later take into account 
this fact using the $^{14}$C+$n$+$n$ model~\cite{horiuchi}. 
We consider two configurations 
for $^{17}$C, the $(0d_{5/2})^2(1s_{1/2})$ and  
$(0d_{5/2})^3$ configurations. 
We assume the ground state spin of $^{19}$C to be 
$\frac{1}{2}^+$ and put the last neutron in the $1s_{1/2}$ orbit,  
following its one-neutron halo structure.  
The protons are assumed to occupy the $0s_{1/2}$ and $0p_{3/2}$ 
orbits for all the carbon isotopes. 

The single-particle orbits arranged according to 
Table~\ref{config} are generated from the following mean-field 
potential 
\begin{equation}
   U(r)=-V_0f(r)+V_1r_0^2\,{\vi \ell}\cdot{\vi s}
   {\frac{1}{r}}\frac{d}{dr}f(r)+V_c(r)\frac{1-\tau_3}{2}, 
\label{opt.pot}
\end{equation}
with $f(r)$=$[1+{\rm exp}\{(r-R) /{a}\}]^{-1}$.  
The radius and diffuseness parameters are chosen as 
$R=r_0A^{1/3}$ with $r_0=1.25\,$fm and $a=0.65\,$fm. 
The spin-orbit strength is set to follow the 
standard value~\cite{bm}, 
\begin{equation}
   V_1=22-14\frac{N-Z}{A}\,\tau_3\ \ \ ({\rm MeV}),
\label{lsstrength}
\end{equation}
whereas the strength $V_0$ of 
the central part for neutron or proton is chosen so as to set 
the binding energy of the last nucleon equal to its 
separation energy, respectively. 
%The single-nucleon asymptotics of the wave function 
The asymptotic form of the single-nucleon wave function 
is satisfied by this requirement, 
which is important for the cross section calculation 
as the surface region determines 
the range of reaction probability.
Table~\ref{av.pot} lists the $V_0$ 
values for both neutron and proton. 
The Coulomb potential for the proton orbits is taken as
\begin{equation}
   V_c(r)=
\begin{cases}
   \frac{(Z-1)e^2}{R}\left[\frac{3}{2}-\frac{1}{2}
   \left(\frac{r}{R}\right)^2\right] & \text{for  $r\le R$} \\
   \frac{(Z-1)e^2}{r}                & \text{for  $r > R$}.
\end{cases}
\end{equation}
For the sake of simplicity, the radius parameter $R$ is assumed to be 
the same as that of the mean-field potential. 

\begin{table}[t]
\caption{Potential parameters $V_0$ in MeV 
in the mean-field model and in the core+$n$ model which is applied to 
the odd $N$ isotope. Two sets are used 
for $^{14}$C: the shallower potential reproduces 
the neutron separation 
energy, while the deeper one is more appropriate to reproduce the 
size of $^{14}$C. Two sets for $^{17}$C correspond to the two 
different configurations in Table~\ref{config}.}
\label{av.pot}
\begin{center}
\begin{tabular}{cccccc}
\hline\hline
        & & \multicolumn{2}{c}{Mean field} & & Core+$n$\\
\hline
Nucleus & & neutron & proton & & neutron  \\
\hline
$^{12}$C & & 57.83 & 57.93 & & \\
$^{13}$C & & 41.99 & 58.42 & & 46.41 \\	
$^{14}$C & & 45.84 & 61.60 & & \\
         & & 53.56 & 61.60 & & \\
$^{15}$C & & 40.09 & 60.34 & & 50.31 \\
$^{16}$C & & 49.28 & 60.99 & & \\
$^{17}$C & & 40.81 & 60.72 & & 44.52 \\
         & & 39.83 & 60.72 & & \\
$^{18}$C & & 46.29 & 63.33 & & \\
$^{19}$C & & 37.84 & 63.59 & & 40.91 \\
$^{20}$C & & 41.27 & 65.04 & & \\	  
\hline
\end{tabular}
\end{center}
\end{table}

The c.m. motion has to be subtracted appropriately 
from the Slater determinant in order to generate the intrinsic 
densities. The neutron or proton intrinsic density is defined as 
\begin{equation}
   \rho({\vi r})=\langle \Psi_0 \mid \sum_{i}\delta(
   {\bar {\vi r}}_i-{\vi X}-{\vi r})P_i \mid \Psi_0 \rangle,
\end{equation}
where ${\bar {\vi r}}_i$ is the single-particle coordinate, 
${\vi X}$ is the c.m. coordinate, and 
$P_i$ is a projector for neutron or proton. 
Denoting the Slater determinant by $\Psi$, we obtain the neutron 
or proton density which contains the effect of the c.m. motion as 
\begin{equation}
   {\widetilde \rho}({\vi r}) = \langle \Psi \mid \sum_{i}\delta(
   {\bar {\vi r}}_i-{\vi r})P_i \mid \Psi \rangle =\sum_{nljm}\mid 
   \psi_{nljm}({\vi r})\mid^2,
\end{equation}
where the sum extends over the occupied neutron or proton  orbits 
depending on $P_i$. 
When the orbit with a certain $nlj$ is not fully occupied, 
the average over $m$ 
is taken in the above summation, that is $\sum_m$ indicates 
$\Omega_j/(2j+1)\sum_{m=-j}^j$ with $\Omega_j$ being the number of
neutrons occupying the $nlj$ orbit. If the Slater determinant 
is approximated as a product of the intrinsic wave function $\Psi_0$ 
and the c.m. part $\Psi_{\rm cm}({\vi X})$, 
\begin{equation}
   \Psi=\Psi_0 \Psi_{\rm cm}({\vi X}),
\label{bmtheorem}
\end{equation}
where 
\begin{equation}
   \Psi_{\rm cm}({\vi X})=\Big(\frac{2A\nu}{\pi}\Big)^{3/4}
   \exp\Big(-A\nu X^2\Big)
\end{equation}
with a suitable oscillator parameter $\nu$, 
it is easy to show that 
\begin{equation}
   \int d{\vi r}{\rm e}^{i{\vis k}\cdot{\vis r}}\rho({\vi r})
   =\exp\Big(\frac{k^2}{8A\nu}\Big)
   \int d{\vi r}{\rm e}^{i{\vis k}\cdot{\vis r}}
   {\widetilde \rho}({\vi r}).
\label{ft.density}
\end{equation}
Since the Fourier transform of $\widetilde{\rho}$ is easily 
obtained, 
the above formula enables us to calculate the intrinsic density 
$\rho$ through an inverse Fourier transformation of the 
right-hand side of Eq.~(\ref{ft.density}).  

The separability of Eq.~(\ref{bmtheorem}) is in general not 
valid, but 
holds exactly for such a case that the Slater determinant is built 
from the lowest configuration of the harmonic-oscillator 
shell model. We test the validity of separability 
by calculating the following overlap
\begin{equation}
   o(\nu)=\frac{1}{A}\sum_{nljm}\mid \langle 
                 \psi_{nljm}^{\rm HO}(\nu)\mid 
   \psi_{nljm}\rangle \mid^2,
\end{equation}
where $\psi_{nljm}^{\rm HO}(\nu)$ is the harmonic-oscillator 
single-particle wave function with the oscillator parameter $\nu$, 
and the sum of $nljm$ 
is taken over both the occupied neutron and proton orbits. 
We search for such $\nu$ that maximizes $o(\nu)$. The values of $\nu$ 
and $o(\nu)$ determined in this way are listed in 
Table~\ref{cm.correction}. We find that $o(\nu)$ is close 
to unity, larger than 0.98 for even $N$ isotopes, so that the 
intrinsic density may be calculated with use of Eq.~(\ref{ft.density}).
The $o(\nu)$ value for odd $N$ nuclei decreases to about 0.95. 
The separability for this case is not as good as for 
even $N$ case, 
but the separability assumption may still be acceptable.

%Table IV
\begin{table}[b]
\caption{Criterion on the separability of the c.m. 
motion from the Slater 
%determinantal 
determinant
wave functions. $\nu$ is the 
parameter of the harmonic-oscillator potential well. See 
Table~\ref{av.pot} for the two sets of $^{14}$C and $^{17}$C.} 
\label{cm.correction}
\begin{center}
\begin{tabular}{ccc}
\hline\hline
Nucleus &$(2\nu)^{-1/2} ({\rm fm})$&$o(\nu)$\\
\hline
$^{12}$C &1.61&0.998\\
$^{13}$C &1.72&0.988\\
$^{14}$C &1.69&0.992\\
  &1.64&0.996\\
$^{15}$C &1.79&0.944\\
$^{16}$C &1.71&0.992\\
$^{17}$C &1.82&0.958\\
  &1.82&0.973\\
$^{18}$C &1.75&0.988\\
$^{19}$C &1.89&0.949\\
$^{20}$C &1.83&0.980\\
\hline\hline
\end{tabular}
\end{center}
\end{table}

\begin{figure}[t]
\epsfig{file=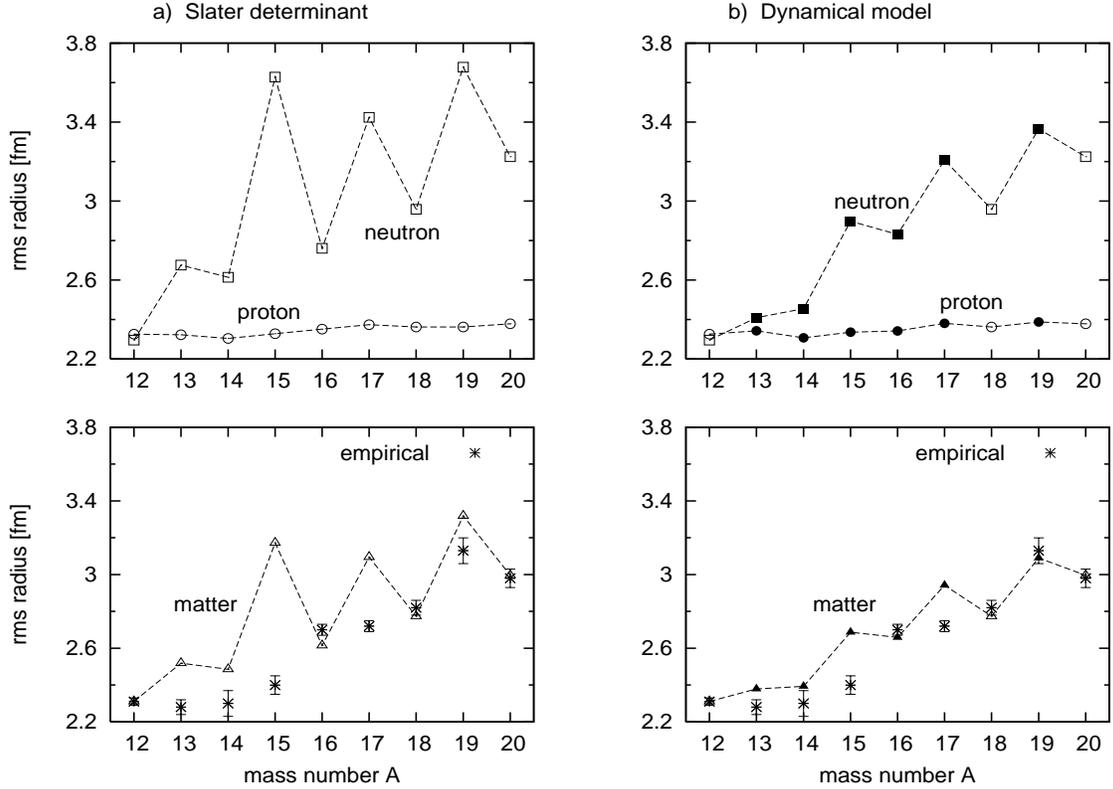,width=15.3cm,height=10.71cm}
\caption{Neutron, proton and matter radii of the carbon isotopes. 
The results with the Slater determinants are shown in panel a), while 
those improved with the core+$n$ and core+$n$+$n$ models are shown 
in panel b). See the text for detail. 
The empirical matter radii are taken from Ref.~\cite{ozawa}. }
\label{radius}
\end{figure}

Figure~\ref{radius}a displays the root mean square (rms) radii 
of neutron, proton and matter distributions assuming a 
point-like nucleon. Corresponding to the large proton separation 
energies of the carbon isotopes, the proton radii remain 
nearly constant in the range 
of 2.3--2.4 fm. Assuming the charge radius of 
the proton to be 0.85 fm, we find that the charge radii of 
$^{12,13,14}$C  
are 2.48, 2.47 and 2.52 fm, respectively, which are 
compared to the experimental values~\cite{ruckstuhl}, 
2.4715, 2.4795 
and 2.4962 fm. The agreement with experiment is very good. 

In contrast to the proton radius, the neutron radii 
change drastically reflecting the even-oddness of the 
neutron number. The isotope with odd $N$  
has much smaller neutron separation energy than the 
isotope with $N-1$. Consequently  
the value of $V_0$ becomes small and all the occupied 
neutron orbits tend to extend to larger distances, resulting in a 
considerable increase of the rms radius. 
See Fig.~\ref{dens.19C} later. 
As the reaction cross section for an odd $N$ nucleus will turn 
out to be too large, we will discuss its density 
beyond the mean-field approach. 

Another point to be noted in Fig.~\ref{radius}a is that 
the neutron radius of $^{14}$C 
given by the present model is much larger 
than that of $^{12}$C. This is in contrast to the 
result of Ref.~\cite{kanada}, in which the radii of both nuclei 
remain almost the same. In fact, we will see later 
that the reaction cross section for $^{14}$C+$^{12}$C is 
too large to be compared to experiment. 
Thus the present mean-field 
description does not seem to work well for $^{14}$C, 
and the molecular model~\cite{itagaki} 
or $^{12}$C+$n$+$n$ three-body model 
may be promising in producing its better density. 
Related to the radius problem 
of $^{14}$C, we note that the $V_0$ value 
for $^{14}$C strongly deviates 
from the systematics of the potential strength for even $N$ nucleus. 
According to Ref.~\cite{bm}, the potential strength for a neutron is 
\begin{equation}
   V_0=51-33\frac{N-Z}{A}\ \ \ ({\rm MeV}).
\label{ls-standard}
\end{equation}
Compared to this value, the $V_0$ value listed in the table 
is deeper by about 6 MeV for even $N$ nuclei, but 
it is nearly equal for $^{14}$C. We test a deeper value of $V_0$ for 
$^{14}$C as listed in Table~\ref{av.pot}. 
This parameter set turns out 
to be more suitable for $^{14}$C, so it will be used to generate 
the densities of $^{15}$C and $^{16}$C in the dynamical model which 
will be discussed in the next section.

\section{Density with a dynamical model}
\label{dens.dynamic}

In the previous section, 
the neutron (proton) separation energy was used to determine 
all of the occupied single-particle orbits. For the nucleus 
with odd $N$, the neutron separation energy is small, so all the 
neutrons result in moving in a shallow potential well. Because of 
this, the radii of the odd $N$ isotopes tend to be too large. 
To improve this restricted description, one has to go beyond a 
Slater determinant model by allowing for the degree of freedom 
of ``clustering''. 
The isotope with odd $N$ will be described with a core nucleus 
with the even number ($N-1$) of neutrons and a neutron. 
Here the last neutron is required to 
have the experimental separation energy, while the core nucleus is 
described as its subsystem 
independently from the separation energy of the last neutron.  

We also consider the partition of a particular 
system into a core nucleus plus 
two neutrons, e.g., a $^{14}$C+$n$+$n$ model for 
$^{16}$C and a $^{20}$C+$n$+$n$ model for $^{22}$C. 
The motivation for this model is as follows. 
The last two neutrons in $^{16}$C are found to have nearly equal 
amount of $(1s_{1/2})^2$ and $(0d_{5/2})^2$ 
configurations~\cite{horiuchi,yamaguchi}. 
It is thus impossible to approximate the ground state of $^{16}$C 
with a single Slater determinant. 
As for $^{22}$C, $^{21}$C is unstable with respect to a neutron 
emission, and $^{22}$C becomes a Borromean system as 
the partition of $^{20}$C+$n$+$n$. 
Thus the core+$n$+$n$ model appears more realistic for $^{22}$C than 
the Slater determinant model. These core+$n$+$n$ models have been 
worked out in Refs.~\cite{suzukic16,horiuchi,c22}.

\subsection{Density in a core+$n$ model}
\label{dens.coren}

A core+$n$ model is applied to the odd isotopes, 
$^{13,15,17,19}$C, where the corresponding 
cores are $^{12,14,16,18}$C, 
respectively. For $^{17}$C, the last neutron is assumed to be 
in the $1s_{1/2}$ orbit. 
Let $\Psi_0=\Phi_{\rm c}\Phi_{1n}$ denote the 
intrinsic wave function of the core+$n$ model, where 
$\Phi_{\rm c}$ represents the intrinsic wave function 
of the core nucleus and $\Phi_{1n}$ the relative motion 
function between the neutron and the core nucleus. The core nucleus 
can be described in exactly the same way as in the previous 
section, while the motion of the 
last neutron for a specified quantum number 
is determined from  the $n$-core 
potential taken as the form of Eq.~(\ref{opt.pot}) 
with $A\, (N)$ being 
replaced by $A-1\, (N-1)$, the mass (neutron) number 
of the core nucleus. 
The potential strength $V_0$ is set to 
reproduce the neutron separation energy, and it is listed in 
Table~\ref{av.pot}. 

% Fig. 4.
\begin{figure}[t]
\epsfig{file=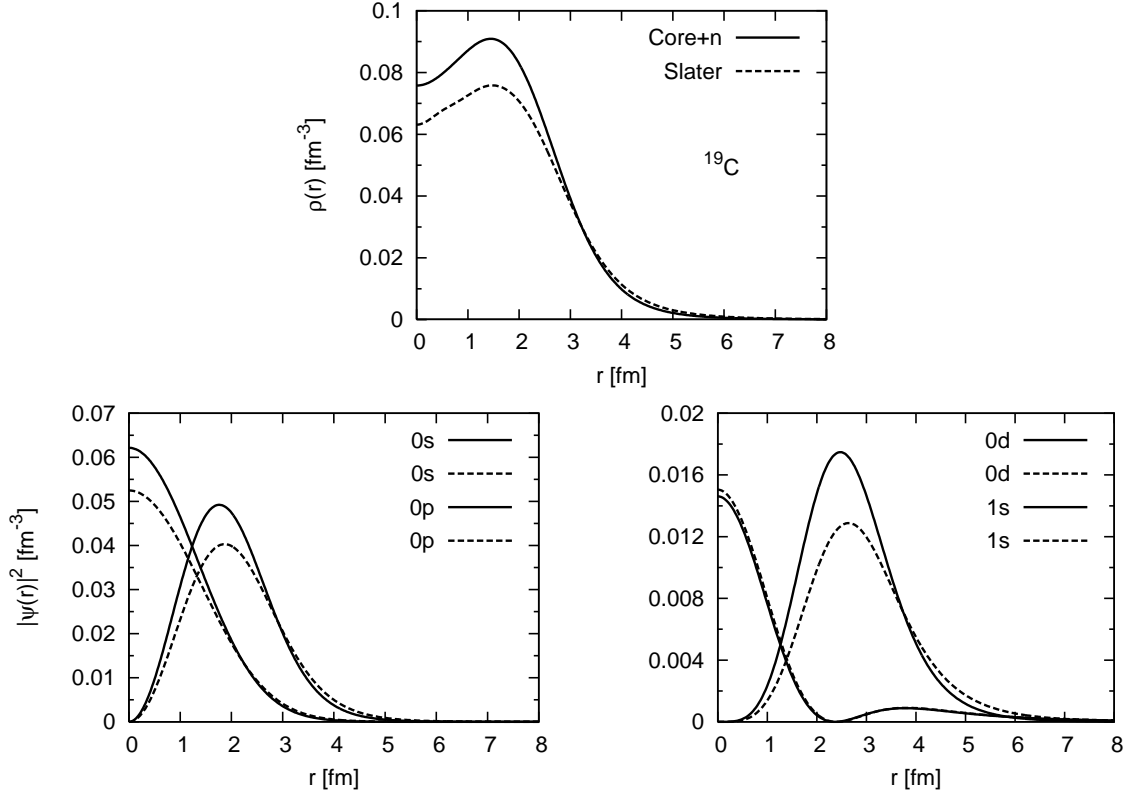,width=15.3cm,height=10.71cm}
\caption{Comparison of the neutron density of $^{19}$C between 
the Slater determinant model and the dynamical model of 
$^{18}$C+$n$. The lower panels show the decomposition of the 
density into the contributions of 
the neutron orbits with different orbital angular momenta. 
The c.m. motion is included. }
\label{dens.19C}
\end{figure}

The intrinsic proton density is given by 
\begin{equation}
   \rho^p({\vi r})=\langle \Phi_{1n}({\vi \rho}) \mid 
   \rho^p_{\rm c}(\textstyle{\frac{1}{A}}{\vi \rho}+{\vi r})\mid 
   \Phi_{1n}({\vi \rho}) \rangle,   
\end{equation}
and similarly the neutron density is
\begin{equation}
   \rho^n({\vi r})=\langle \Phi_{1n}({\vi \rho}) \mid 
   \rho^n_{\rm c}(\textstyle{\frac{1}{A}}{\vi \rho}+{\vi r})\mid 
   \Phi_{1n}({\vi \rho}) \rangle + \langle  \Phi_{1n}({\vi \rho})
   \mid \delta(\textstyle{\frac{A-1}{A}}{\vi \rho}-{\vi r}) 
   \mid  \Phi_{1n}({\vi \rho}) \rangle.
\end{equation}
Here ${\vi \rho}$ is the distance vector from the c.m. of the core 
to the last neutron, and 
$\rho_{\rm c}$ is the intrinsic density of the core nucleus. 
The integration with respect to the spin coordinate of the neutron 
should be done though it is not explicitly written in these 
equations. 

We compare in Fig.~\ref{dens.19C} 
the neutron density of $^{19}$C between the Slater 
determinant model and the dynamical $^{18}$C+$n$ model. 
The contribution of the neutron orbits 
to the density is also displayed in two lower panels 
for each orbital angular momentum 
in order to clarify the difference in the radial extension. 
The density of the Slater determinant model extends radially 
to further distances than that of the dynamical model. 
Particularly the $0p$ and $0d$ orbits play the most important 
role in producing the different neutron size.   

The resulting radii calculated in the core+$n$ model are 
displayed in Fig.~\ref{radius}b by closed square (for neutron), 
closed circle (for proton) and closed triangle (for matter), 
respectively. We see that the core+$n$ model 
leads to a substantial reduction in the neutron radius, resulting 
in a fair improvement for the matter radius. The matter 
radius of $^{19}$C especially is in good agreement with the 
empirical value. The matter radii of $^{15}$C and $^{17}$C are, 
however, still too large compared to the empirical ones.

\subsection{$^{22}$C in a $^{20}$C+$n$+$n$ model} 
\label{carbon22}

The ground state of $^{22}$C in the core+$n$+$n$ model~\cite{c22} 
is given by 
\begin{equation}
   \Psi=\Phi_{\rm c}\Phi_{2n}({\vi x}_1,{\vi x}_2),
\label{totalwf}
\end{equation}
where the two-neutron wave function $\Phi_{2n}$ is expressed with 
the $n$-core relative coordinates, ${\vi x}_1$ and ${\vi x}_2$, 
again suppressing the spin coordinates. 
The valence neutron part $\Phi_{2n}$ is obtained in  
a combination of correlated Gaussian bases, $\Phi_{2n}=\sum_i 
C_i\Phi(\Lambda_i,A_i)$, with  
\begin{equation}
   \Phi(\Lambda,A)\!=\!(1-P_{12})  \left\{ {\rm e}^{-{\frac{1}{2}}
   \tilde{\vis x}A{\vis x}} [[{\cal Y}_{\ell}({\vi x}_1)
   {\cal Y}_{\ell}({\vi x}_2)]_L\chi_S(1,2)]_{00}\right\},
\label{base}
\end{equation}
where $P_{12}$ permutes the neutron coordinates and 
$\tilde{\vi x}A{\vi x}= 
A_{11}{\vi x}_1^{\, 2}$+$2A_{12}{\vi x}_1
\!\cdot\!{\vi x}_2$+$A_{22}{\vi x}_2^{\, 2}$. 
The angular parts of the two-neutron motion are described using 
${\cal Y}_{\ell m}({\vi r})\!=\!r^{\ell}Y_{\ell m}(\hat{\vi r})$ and 
they are coupled with the spin part $\chi_S$ to the total angular 
momentum zero. The basis function is specified by a set of angular 
momenta $\Lambda$=$(\ell,S)$ ($L$=$S$), and a 2$\times$2 
symmetric matrix $A$ ($A_{21}$=$A_{12}$). 
The two neutrons are explicitly correlated due to the term 
$A_{12}{\vi x}_1$$\cdot$${\vi x}_2$, the inclusion of which 
assures a precise 
solution in a relatively small dimension~\cite{svm}. 

The two-neutron wave function $\Phi_{2n}$ 
is determined by solving the 
relevant three-body problem of the Hamiltonian
\begin{equation}
   H=T_{\vis \lambda}+T_{\vis \rho}+U_1+U_2+v_{12}
\label{hamiltonian}
\end{equation}
under the Pauli constraint that $\Phi_{2n}$ should be orthogonal 
to any orbits occupied in the core nucleus $^{20}$C.   
The subscripts, ${\vi \lambda}$ and ${\vi \rho}$, of the 
kinetic energies stand for 
the relative distance vectors of the three-body system: 
${\vi \lambda}={\vi x}_1-{\vi x}_2$ and ${\vi \rho}=
\frac{1}{2}({\vi x}_1+{\vi x}_2)$. 
The two-neutron potential $v_{12}$ 
is taken from the realistic G3RS (case 1) potential~\cite{tamagaki} 
which contains central, tensor and spin-orbit forces 
and reproduces the nucleon-nucleon scattering data 
as well as the deuteron properties. The 
$n$-$^{20}$C potential $U_i$ is taken in the form 
of Eq.~(\ref{opt.pot}) augmented with an additional term: 
\begin{equation}
   U=-V_0f(r)+V_1r_0^2{\vi \ell}\cdot{\vi s}{\frac{1}{r}}
   {\frac{d}{dr}}f(r)+
    V_s{\rm e}^{-\mu r^2}{\cal P}_s.
\end{equation}
The operator ${\cal P}_s$ of the last term projects to the $s$ 
wave of the $n$-$^{20}$C relative motion, so this term 
modifies the $s$-wave potential strength. In evaluating 
%the Hamiltonian matrix element 
angular-momentum dependent matrix elements 
in the basis of Eq.~(\ref{base}), we have 
neglected a small difference between the ${\vi x}_1,\, {\vi x}_2$ 
coordinate and the Jacobi coordinate as the core mass is 
much larger than the neutron mass. 
To determine the parameters of $U$, we 
take into account the conditions that (i) the $1s_{1/2}$ orbit is 
unbound as $^{21}$C is unstable for a neutron emission, 
and (ii) the $0d_{5/2}$ orbit is bound by at most 2.93 MeV, 
which is the neutron separation energy of $^{20}$C. Since 
no information is available to 
determine the $s$-wave strength except that the $1s_{1/2}$ orbit is 
unbound, we vary $V_s$ in a reasonable range. The range parameter 
$\mu$ is set to be $\mu=0.09\,$fm$^{-2}$. The value of $V_0$ 
is 43.24 MeV (set B of Ref.~\cite{c22}) and $V_1$ is 
fixed to be 25.63 MeV ($N=14,\, Z=6$ in Eq.~(\ref{lsstrength})). 

Table~\ref{c22-scat.length} lists the ground state energy $E$ of 
$^{22}$C with respect to the $^{20}$C+$n$+$n$ threshold together with 
the rms neutron, proton and matter radii for some values of $V_s$. 
The calculated energies are all within the uncertainty of the 
experimental value ($-$0.423$\pm$1.140 MeV). If one chooses a 
smaller value than 9.46 MeV for $V_s$, the $1s_{1/2}$ orbit 
would be bound. We see from the table that the neutron radius 
increases considerably as the $s$ wave potential strength decreases. 
A slight change of the proton radius is due to the change of the 
two-neutron wave function, as will be discussed in the next 
subsection.  

% Table 5
\begin{table}[t]
\caption{Properties of $^{22}$C for different $V_s$ values of the 
$n$-$^{20}$C potential. $E$ is the ground state energy in MeV 
with respect to the $^{20}$C+$n$+$n$ threshold, and 
$\langle r^2_{\, n}  \rangle^{1/2}$, $ \langle r^2_{\, p}
 \rangle^{1/2}$ and $\langle r^2_{\, m} \rangle^{1/2}$ denote the rms 
neutron, proton and matter radii given in fm, respectively.} 
\label{c22-scat.length}
\begin{center}
\begin{tabular}{ccccc}
\hline\hline
$V_s$ & $E$ & $\langle r^2_{\, n}  \rangle^{1/2}$ & 
$ \langle r^2_{\, p}  \rangle^{1/2}$ & 
$\langle r^2_{\, m} \rangle^{1/2}$ \\
\hline
9.46&$-$0.489& 3.96 & 2.43 & 3.61\\
9.90&$-$0.361&4.07&2.44&3.69\\
10.4&$-$0.232&4.24&2.45&3.83\\
10.9&$-$0.122&4.58&2.48&4.11 \\
\hline\hline
\end{tabular}
\end{center}
\end{table}

\subsection{Density in a core+$n$+$n$ model}
\label{dens.corenn}

The intrinsic neutron density for the core+$n$+$n$ system 
is obtained by  
\begin{equation}
   \rho^{n}({\vi r})=\langle \Phi_{2n}({\vi x}_1,{\vi x}_2) 
   \mid \rho_{\rm c}^{n}(\textstyle{\frac{2}{A}}{\vi \rho}
   +{\vi r})\mid \Phi_{2n}({\vi x}_1,{\vi x}_2) \rangle
   + \rho_{2n}({\vi r}), 
\label{dens.n.corenn}
\end{equation}
where 
\begin{equation}
   \rho_{2n}({\vi r})=\langle \Phi_{2n}({\vi x}_1,{\vi x}_2) 
   \mid \sum_{i=1}^2
   \delta({\vi x}_i-\textstyle{\frac{2}{A}}{\vi \rho}-{\vi r})
   \mid \Phi_{2n}({\vi x}_1,{\vi x}_2) \rangle
\label{dens.2n.corenn}
\end{equation}
is the contribution of the two neutrons to the neutron density. 
The intrinsic proton density is given by 
\begin{equation}
   \rho^{p}({\vi r})=\langle \Phi_{2n}({\vi x}_1,{\vi x}_2) \mid 
   \rho_{\rm c}^{p}(\textstyle{\frac{2}{A}}{\vi \rho}
   +{\vi r})\mid \Phi_{2n}({\vi x}_1,{\vi x}_2) \rangle. 
\label{dens.p.corenn}
\end{equation}
We use the intrinsic core density obtained 
in Sect.~\ref{dens.slater}. 
A method of calculation for the density with the correlated Gaussians 
$\Phi_{2n}({\vi x}_1,{\vi x}_2)$ is given in Appendix. 

We compare in Fig.~\ref{dens.16C} the densities for $^{16}$C 
obtained with the Slater determinant and the 
dynamical core+$n$+$n$ model. The dynamical model with 
$^{14}$C+$n$+$n$ allows us to include both of 
the $d$ and $s$ waves for the last neutrons. This is the reason 
why the central density rises compared to that with the 
Slater determinant where the last two neutrons are restricted to 
the $(0d_{5/2})^2$ configuration. It is also noted that the 
density of the dynamical model is larger at large distances 
($r \ge 4.0$ fm) 
than that of the Slater determinant model. 
As shown in Fig.~\ref{radius}b, 
the matter radius of 
$^{16}$C calculated in the dynamical model slightly increases 
compared to that of the Slater determinant model, and it is in 
good agreement with the empirical value.

% Fig. 5.
\begin{figure}[b]
\epsfig{file=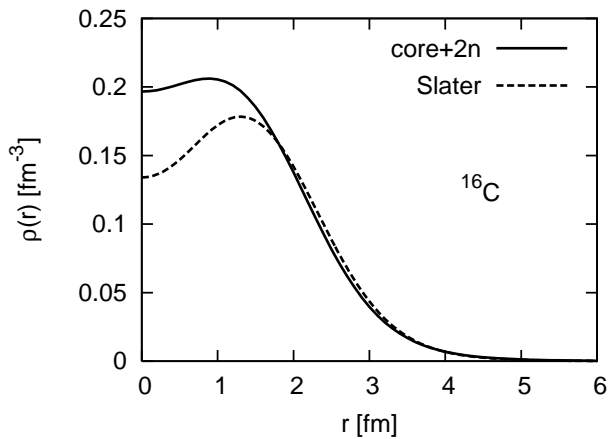,width=8.5cm,height=5.95cm}
\caption{Comparison of the matter density of $^{16}$C between the 
Slater determinant model and the $^{14}$C+$n$+$n$ model. }
\label{dens.16C}
\end{figure}

\begin{figure}[t]
\epsfig{file=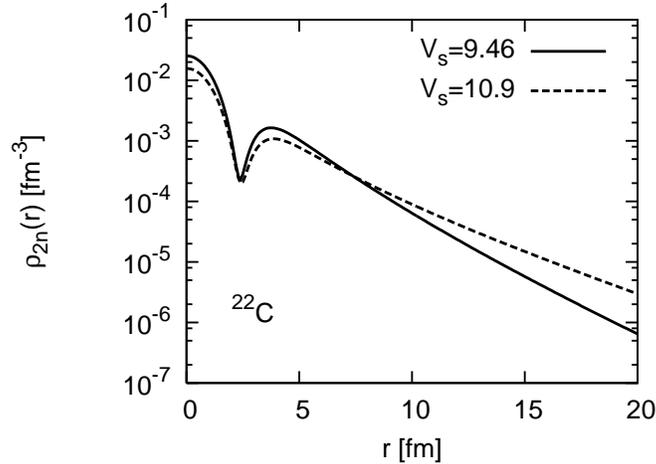,width=9.01cm,height=6.307cm}
\caption{The two-neutron densities of $^{22}$C for some of the 
potential parameters of Table~\ref{c22-scat.length}. }
\label{dens.22C}
\end{figure}

Figure~\ref{dens.22C} displays the two-neutron density distribution 
$\rho_{2n}({\vi r})$ of $^{22}$C for the potential parameters 
given in 
Table~\ref{c22-scat.length}. The density decreases slowly for 
increasing $r$, reaching 
far distances. The two-neutron density is found to dominate the total 
neutron density of $^{22}$C for $r > 6$ fm~\cite{c22}. 
The position of the dip hardly alters against the change of 
$V_s$, which is because the dip appears as a consequence of 
the Pauli orthogonality constraint to the orbits occupied in the core 
mentioned above.  

% Fig. 6.

\section{Numerical Results}
\label{result}

In this section, we show our numerical results 
of the total reaction cross sections of 
the carbon isotopes on $^{12}$C.

\subsection{Reaction cross sections for $^{12}$C to $^{20}$C}
\label{rcs.c12-20}

%Fig.7
\begin{figure}[b]
\epsfig{file=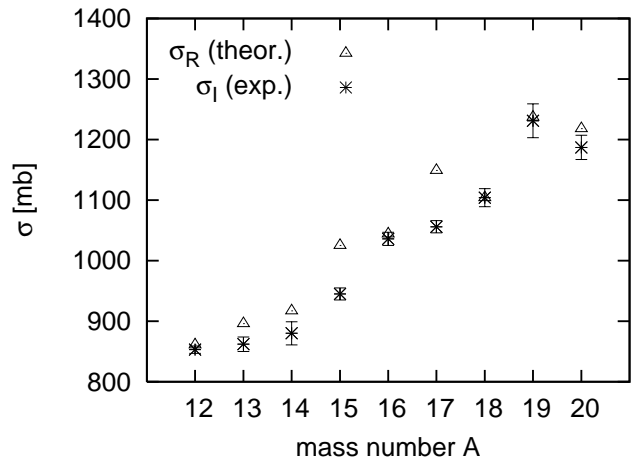,width=9.01cm,height=6.307cm}
\caption{
The reaction cross sections of the carbon isotopes 
on a $^{12}$C target calculated with the NTG model. 
The experimental data cited here are the 
interaction cross sections~\cite{ozawa}. 
The incident energy of the projectile nucleus 
is around 950 $A\,$MeV, except 
for $^{15}$C performed at 740 $A\,$MeV.}
\label{HErcs.ciso-c12}
\end{figure}

The calculation of reaction cross section has been performed 
using the phase shift functions defined by Eqs.~(\ref{ola}) (OLA), 
(\ref{gptsym}) (NTG) and (\ref{gntola}) together with Eq.~(\ref{eik}) 
(NTO). The densities used in the reaction calculation 
give the nucleon radii shown in panel b) of Fig.~\ref{radius}. 
These densities are fitted in terms of a combination of Gaussians 
with different width parameters to facilitate the phase shift 
calculation. 

In Fig.~\ref{HErcs.ciso-c12}, 
we plot the numerical results of NTG calculation of
the reaction cross sections of the carbon isotopes on $^{12}$C at 
high incident energy around 950 $A\,$MeV. 
For comparison, we plot the data in the same figure. 
To compare the numerical results with experiment, 
we have to bear in mind that 
most of the cross sections measured at high energy are 
not reaction cross sections but interaction cross sections. 
The interaction cross section does not include 
the contribution from those 
inelastic processes which correspond to the excitation of the 
projectile to particle-bound excited states, 
so the interaction cross section is in general smaller 
than the total reaction cross section if such excited states exist. 
Actually, as was pointed out recently~\cite{KIO}, 
there exists some difference between the total reaction cross 
section and the interaction cross section, about 80 mb, 
for the case of $^{12}$C+$^{12}$C reaction. 
Since such difference depends on the nuclear structure, 
and no such data is available, 
we consider the difference as a kind of maximum 
uncertainty of our numerical results in this figure.

A comparison with experiment indicates that 
the numerical results of 
the reaction cross sections for $^{12,16,18,19,20}$C 
agree with the interaction cross section data, 
whereas those for $^{15,17}$C are too large.

Whether the reaction cross section calculated for $^{14}$C is 
quite reasonable or a 
little too large compared to experiment is not clear, 
because it is difficult to estimate possible 
contribution from the inelastic processes. 
For all the carbon isotopes with even $N$ (except for $^{14}$C), 
we have the 
densities which reproduce the experimental cross sections. We used 
these densities in the core+$n$ description for the carbon isotopes 
with odd $N$. The reaction cross sections calculated with this 
model is found to bring a significant improvement in the 
agreement with experiment.  
Particularly, the agreement attained in $^{19}$C 
is excellent, considering that the reaction cross section is 
equal to the interaction cross section for $^{19}$C to good accuracy. 
For the case of $^{13}$C, it is not clear whether the 
differences between theory and experiment can entirely 
be explained by 
the difference between the reaction cross section and the 
interaction cross section. 

The reaction cross section for $^{16}$C is calculated in the 
$^{14}$C+$n$+$n$ model using the improved density of $^{14}$C. 
The two neutrons are restricted to 
neither $(1s_{1/2})^2$ nor $(0d_{5/2})^2$ configuration, but 
contains both of them together with other 
configurations~\cite{horiuchi}.  
We see that the calculated cross section 
turns out to be in almost perfect agreement with experiment 
within its error. 

\begin{figure}[t]
\epsfig{file=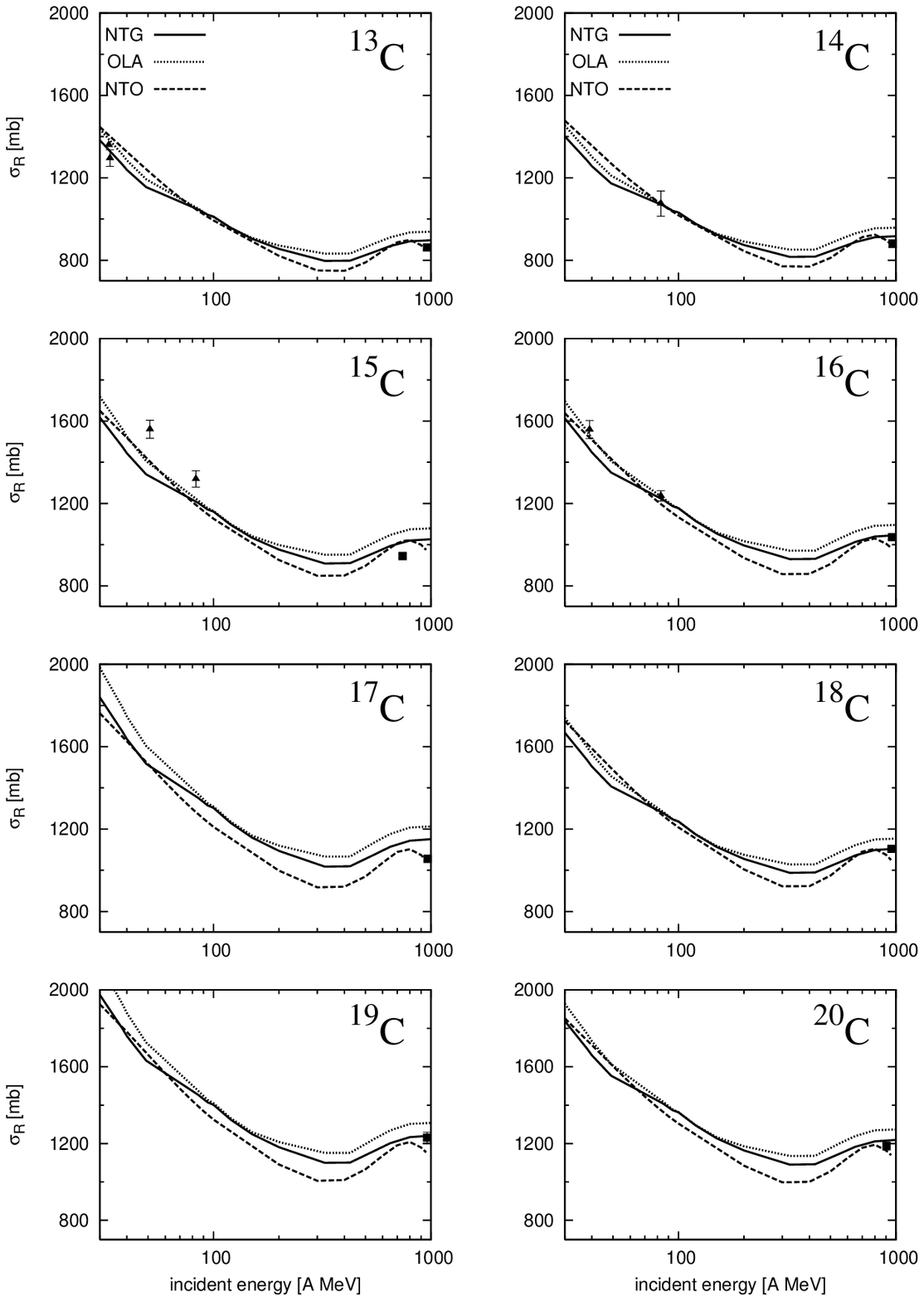,width=13cm,height=18.2cm}
\caption{Reaction cross sections for the collisions of the 
carbon isotopes on a $^{12}$C target calculated with the NTG, 
NTO and OLA models. The experimental data marked with the closed 
triangle and the closed square denote the reaction cross 
section and the interaction cross section~\cite{ozawa}, 
respectively. The reaction cross section data are taken from 
Refs.~\cite{fukuda91,fang00} for $^{13}$C, from Ref.~\cite{fang} 
for $^{14,15}$C and from Refs.~\cite{fang00,zheng} for $^{16}$C.}
\label{ciso-c12.reac}
\end{figure}

We predict in Fig.~\ref{ciso-c12.reac} the 
reaction cross sections of the carbon isotopes $^{13-20}$C 
on a $^{12}$C target as a function of the incident energy. 
The reaction cross section predicted by the OLA 
is typically 50 mb larger than that predicted by the NTG 
except for the incident energy range of 80--150 $A\,$MeV. This 
tendency is already seen in the $^{12}$C+$^{12}$C case, as 
shown in Fig.~\ref{c12-c12reac}. The energy dependence 
of both NTG and NTO cross sections is similar to that 
of the $^{12}$C+$^{12}$C case displayed in Fig.~\ref{c12-c12reac}. 
Very limited experimental data available at lower 
energies hamper a clear-cut conclusion. It appears, however, that 
the cases for $^{13,14,16}$C are successfully reproduced. 
In contrast to these nuclei, the cross section of $^{15}$C 
clearly indicates a marked discrepancy between theory and 
experiment: The theory underestimates the cross section at 
lower energy, but appears to overestimate it at high energy. 

One might think that the above discrepancy in $^{15}$C 
could be resolved 
by including its breakup effect into $^{14}$C+$n$ continuum states 
in the few-body (FB) 
framework of a core+$n$ model~\cite{yos,surrey,cpc}. This is not 
the case, however, because the NTO already takes into account 
most of the effect.  
In fact, we have compared the reaction cross sections between 
NTO and FB at several incident energies, and found that the 
difference between them is small even at low energy: 
For example, at the incident energy of 40 $A\,$MeV, 
the $\sigma_{\rm R}$ value of $^{15}$C+$^{12}$C 
is 1519 mb for NTO and 1525 mb for 
FB, whereas it is only 1425 mb for the folding model which 
uses Eq.~(\ref{foldig.psf}) to obtain the phase shift function. 
Thus the increase of the reaction cross section given by 
FB compared to NTO is just 6 mb for $^{15}$C, and 18 mb for $^{19}$C. 
At the higher energy of 800 $A\,$MeV, the FB cross section becomes 
only slightly smaller than the NTO cross section. 
The discrepancy observed in the reaction cross 
section of $^{15}$C+$^{12}$C remains an open question.

\subsection{Reaction cross section for $^{22}$C}
\label{rcs.c22}

\begin{figure}[t]
\epsfig{file=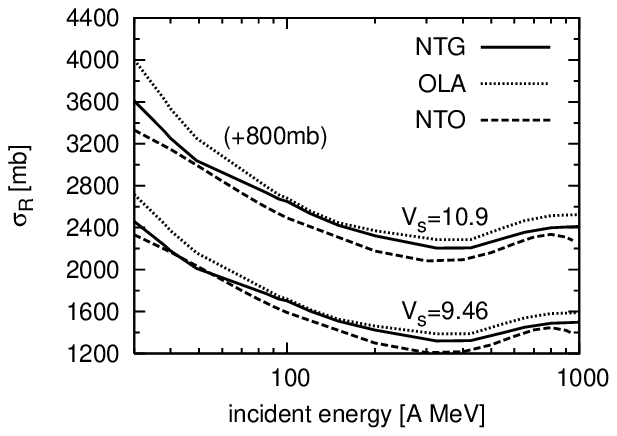,width=8.5cm,height=5.95cm}
\caption{Prediction of the reaction cross section 
for the collision of $^{22}$C 
on a $^{12}$C target as a function of the incident energy. 
See Table V for the two values of $V_s$.}
\label{rcs.22C-12C}
\end{figure}

We display in Fig.~\ref{rcs.22C-12C} our prediction of the 
reaction cross section of $^{22}$C+$^{12}$C as 
a function of the incident energy. As the reaction cross section 
increases for the increasing radius of the projectile and 
no information on the mass of $^{22}$C is available, we plot 
two cases obtained using the densities which correspond to the 
two extremes in Table~\ref{c22-scat.length}, namely $V_s=9.46$ 
and 10.9 MeV. The matter radius 
obtained with the latter parameter is larger by 11 \% than 
that with the former parameter. 

The cross sections calculated 
with the NTG and NTO models are in reasonable agreement, while 
the OLA cross section at lower energy shows an enhancement 
of about 10 \% compared to the NTG cross section. According to 
the NTG calculation, the total reaction cross section of $^{22}$C 
is estimated to be 2200--2450 mb at 40 $A\,$MeV, and 1500--1600 mb 
around 900 $A\,$MeV. 

In order to see the implication of these results,
we refer to the black-sphere picture~\cite{BS1,BS2} 
or the strong absorption model~\cite{slyv}: 
These pictures include only one scale, the nuclear radius, $a$. 
If one determine the values of $a$ so as to reproduce the angle 
of the first diffraction maximum in the proton-nucleus 
elastic scattering data, 
the absorption cross section, $\pi a^2$, agrees 
with the empirical total reaction cross section~\cite{BS2}. 
This $a$ can be regarded as a ``reaction radius",  
inside which the reaction with incident protons occurs.

Since the data of $p$+$^{22}$C elastic differential 
cross section are not available, 
we may estimate the reaction radius through 
\begin{equation}   
   \sigma_{\rm R}({\rm P}+{\rm T})=\pi (R_{\rm P} + R_{\rm T})^2,
\end{equation}
where $R_{\rm P}$ and $R_{\rm T}$ are the reaction radii 
of the projectile and target, respectively~\cite{BS2}. 
For a $^{12}$C target we obtain $R_{\rm T}=2.69$ fm~\cite{BS2}.
The reaction radius of {\em stable} nuclei
follows $1.21 A^{1/3}$ fm. 
If we apply it to the above expression, 
we obtain 1170 mb, much smaller than the result 
around 900 $A\,$MeV.
This supports the much more extended matter distribution 
of $^{22}$C than the stable nuclei of the same mass number.

Based on the above expression, 
the reaction radius of $^{22}$C may be estimated by 
\begin{equation}
   R_{^{22}{\rm C}}
  =\left(\sqrt{\frac{4\sigma_{\rm R}(^{22}{\rm C} + {^{12}{\rm C}})}
          {\sigma_{\rm R}(^{12}{\rm C} + {^{12}{\rm C}})}} - 1 \right)
   R_{^{12}{\rm C}}.
\end{equation}
Using the reaction cross sections calculated 
at high energy together with 
$R_{^{12}{\rm C}}
=\sqrt{\sigma_{\rm R}(^{12}{\rm C}+{^{12}{\rm C}})/{4\pi}}$ 
leads to the estimation 
that the reaction radius of $^{22}$C, 
$R_{^{22}{\rm C}}$, is about 4.41 fm. 
Multiplying it by $\sqrt{3/5}$, we obtain 3.41 fm 
for the rms reaction radius of $^{22}$C, 
which is smaller than the rms matter radii
listed in Table~\ref{c22-scat.length}.
For lighter stable nuclei, typically $A<50$, 
the rms reaction radii are usually smaller than 
the rms matter radii \cite{BS2}. 
Since light nuclei have no sharp surface, 
the reaction occurs inside compared to heavier nuclei. 

Actually, for the data of incident energies 
higher than $\sim800$ $A\,$MeV,
$\sqrt{3/5}a$ systematically 
deviates from the empirically deduced values 
of the rms matter radius
for nuclei having mass number less than about 50, 
while it almost completely agrees with the deduced values for 
$A\gtrsim50$.  
This tendency suggests a significant change of the nuclear 
matter distribution from a rectangular one for $A\lesssim50$, 
which is consistent with the behavior of 
the empirical charge distribution.
Therefore, the above result of $^{22}$C suggests that such feature 
of light nuclei still persists 
in $^{22}$C although it has a large radius comparable to 
much heavier nuclei. 

Measurements of the reaction cross section as well 
as the mass of $^{22}$C 
are indispensable for the determination of its radius.

%Fig.9

\section{Summary}
\label{summary}

We have systematically analyzed the total reaction cross sections 
of carbon isotopes with $N=6$--$16$ on a $^{12}$C target 
for wide range of incident energy from 40 to 1000 $A\,$MeV. 

The structure of the carbon isotopes has first been described 
by a Slater determinant generated from a phenomenological 
mean-field potential. The potential depth of Woods-Saxon type 
is determined separately for 
neutron and proton to reproduce the nucleon 
separation energy. 
The intrinsic density of each carbon isotope is 
built from the single-particle states by separating, 
in a good approximation, 
the center of mass motion from the Slater determinant.  
This model reasonably well describes the ground states 
of even $N$ isotopes,  
but the mean-field potential for odd $N$ isotope tends to be too 
shallow, yielding too large neutron and matter radii.  
This unrealistic 
feature has been largely improved by performing separate 
studies which 
take into account their specific structure of core+$n$.  
We have also 
performed the core+$n$+$n$ three-body model 
for $^{16}$C and $^{22}$C, 
in order to take into account the mixing of the $sd$ orbits 
in $^{16}$C 
and a Borromean character of $^{22}$C, respectively.  

For calculations of the cross sections, 
we take two schemes: one is the Glauber approximation, and 
the other is the eikonal model using a global optical potential. 
It is vital to find a consistent parameterization 
of the nucleon-nucleon scattering amplitude in the former model. 
The parameters we find for the $NN$ profile function are different 
from previous ones, and they have successfully reproduced 
the data on 
$^{12}$C+$^{12}$C reaction cross sections from  40 to 1000 $A\,$MeV 
incident energies. The both reaction schemes reasonably well 
reproduce the data of the cross sections of $^{13}$C, 
$^{14}$C and $^{16}$C on $^{12}$C 
which are available at low and high incident energies. 
Those data which are 
available for $^{18}$C, $^{19}$C and $^{20}$C around 950 $A\,$MeV 
are all reproduced very well by the Glauber theory. 

Compared to the empirical radii of the 
carbon isotopes tabulated in Ref.~\cite{ozawa}, 
our dynamical model 
gives too large values for $^{15}$C (see Fig.~\ref{radius}b). From 
this comparison, we expect that the reaction cross section 
predicted by the present model is larger than the experiment. 
In fact, this is 
true for the high energy data at 740 $A\,$MeV, 
but it is just opposite at 
low energy. See Fig.~\ref{ciso-c12.reac}. 
It is an open question that 
the calculated reaction cross sections of $^{15}$C, 
though our calculation practically includes the breakup effect, 
is found to underestimate the empirical values observed 
at low energy. 

The radius of $^{17}$C is also calculated to be too large. 
Even in the $^{16}$C+$n$ dynamical version, 
we do not consider that the model for $^{17}$C 
is probably very realistic. 
More sophisticated structure model will be necessary. 

We have predicted the total reaction cross section 
of $^{22}$C on $^{12}$C. 
In our model $^{22}$C has extremely large size 
comparable to that of 
a medium heavy nucleus. Our prediction for the cross section is 
at variance 
with the binding energy of the two neutrons:  
According to the Glauber calculation, 
the reaction cross section of $^{22}$C 
is 2200--2450 mb at 40 $A\,$MeV, and 1500--1600 mb 
around 900 $A\,$MeV. 
Measurements of the reaction cross section as well 
as the mass of $^{22}$C will be useful to determine 
the structure of  $^{22}$C. 

Our framework offers a prescription for simple, 
consistent analyses 
of broad range of reaction cross section data 
of neutron-rich unstable nuclei. 
Such data are expected to be provided by radioactive ion beam 
facilities, such as GSI and Radioactive Ion Beam Factory at RIKEN.

\vspace{10mm}

We acknowledge T. Motobayashi for his encouragement
during the course of this work. 
We also acknowledge M. Takechi and M. Fukuda for 
providing us with data on $^{12}$C+$^{12}$C reaction cross sections. 
A.K. would like to thank K. Iida, K. Oyamatsu, and M. Takashina for 
useful comments and helpful discussions. 
This work was in part supported by a Grant for Promotion of Niigata 
University Research Projects (2005--2007).

\appendix
\section*{Appendix: Calculation of 
a two-particle distribution function}

The aim of this appendix is to outline a method of calculation 
for the density which appears in Sect.~\ref{dens.corenn}. 
Expressing the core density in Eqs.~(\ref{dens.n.corenn}) 
and (\ref{dens.p.corenn}) as 
\begin{equation}
\rho_{\rm c}\Big(\frac{2}{A}{\vi \rho}\Big)=\int 
\delta\Big(\frac{2}{A}{\vi \rho}-{\vi r}\Big)\rho_{\rm c}({\vi r}) 
\, d{\vi r}
\end{equation}
with ${\vi \rho}=\frac{1}{2}({\vi x}_1+{\vi x}_2)$, 
we note that the terms in Eqs.~(\ref{dens.n.corenn}), 
(\ref{dens.2n.corenn}) and (\ref{dens.p.corenn}) are all reduced 
to the calculation of the two-particle distribution function
\begin{equation}
D(w,{\vi r})=\langle [G_{S}(A',\ell')\chi_S(1,2)]_{00}
\mid \delta(\tilde{w}{\vi x}-{\vi r}) \mid 
[G_{S}(A,\ell)\chi_S(1,2)]_{00} \rangle.
\end{equation}
Here the function $G$ is a short-hand notation for 
\begin{equation}
G_{LM}(A,\ell)={\rm e}^{-{\frac{1}{2}}
\tilde{\vis x}A{\vis x}} [{\cal Y}_{\ell}({\vi x}_1)
{\cal Y}_{\ell}({\vi x}_2)]_{LM},
\end{equation}
and $\tilde{w}{\vi x}$ stands for 
$w_1{\vi x}_1+w_2{\vi x}_2$, where  
$w_1$ and $w_2$ are constants which are chosen appropriately 
depending on the two-particle distribution function to be evaluated. 
A choice of $w_1$=$1-\frac{1}{A}$ and $w_2$=
$-\frac{1}{A}$ or $w_1$=$-\frac{1}{A}$ and $w_2$=$1-\frac{1}{A}$ 
is made for the evaluation of the density of 
Eq.~(\ref{dens.2n.corenn}), while $w_1$=$w_2$=$\frac{1}{A}$ is 
chosen for Eqs.~(\ref{dens.n.corenn}) 
and (\ref{dens.p.corenn}).  

After integrating over the spin coordinates, we obtain
\begin{equation}
D(w,{\vi r})
=\sum_{\lambda}C_{\lambda}(\ell \ell' S)\int \int {\rm e}^{-{\frac{1}{2}}
\tilde{\vis x}B{\vis x}}(x_1x_2)^{\ell+\ell'}
[Y_{\lambda}(\widehat{{\vi x}_1})Y_{\lambda}(\widehat{{\vi x}_2})]_{00}
\, \delta(\tilde{w}{\vi x}-{\vi r})\, d{\vi x}_1d{\vi x}_2,
\label{app3}
\end{equation}
where $B=A+A'$ and 
\begin{equation}
C_{\lambda}(\ell \ell' S)=\frac{(2\ell+1)(2\ell'+1)}
{4\pi(2\lambda+1)\sqrt{2S+1}}\langle \ell 0 \ell' 0\mid \lambda 0
\rangle^2 U(\ell \lambda S \ell';\ell' \ell ).
\end{equation}
Here $U$ is a unitary Racah coefficient, and $\lambda$ takes 
those values from $\mid \ell-\ell'\mid$ to $\ell+\ell'$ which 
satisfy the condition of $\lambda+\ell+\ell'$=even. 

The integration ${\cal I}$ in Eq.~(\ref{app3}) can be performed 
by a change of 
variables from ${\vi x}$ to ${\vi y}$, ${\vi x}=W{\vi y}$, under the 
condition that ${\vi y}_2$ 
is set equal to $\tilde{w}{\vi x}$. Though ${\vi y}_1$ can be chosen 
arbitrarily as long as it is independent of ${\vi y}_2$, we here 
choose $W$ as follows: 
\begin{equation}
W=\frac{1}{w_1^2+w_2^2}
\begin{pmatrix}
w_2 & w_1 \\
-w_1 & w_2 \\
\end{pmatrix}.
\end{equation}
Substituting ${\vi x}=W{\vi y}$ in Eq.~(\ref{app3}) and noting 
that $[Y_{\lambda}(\widehat{{\vi x}_1})Y_{\lambda}(\widehat{{\vi x}_2})]_{00}$ 
can be expressed in terms of a Legendre polynomial $P_{\lambda}(\zeta)$ with 
$\zeta=({\vi x}_1\cdot{\vi x}_2)/
(x_1x_2)$,   
we obtain
\begin{equation}
{\cal I}
=({\rm det}W)^{3}\int \int {\rm e}^{-{\frac{1}{2}}
\tilde{\vis y}\bar{B}{\vis y}}F_1({\vi y})F_2({\vi y})\, 
\delta({\vi y}_2-{\vi r})\, d{\vi y}_1d{\vi y}_2,
\end{equation}
where $\bar{B}=\widetilde{W}BW$ and 
\begin{eqnarray}
F_1({\vi y})&=& (x_1x_2)^{\ell+\ell'-\lambda}
\nonumber \\
&=&({\rm det}W)^{2(\ell+\ell'-\lambda)}
\Big\{\mid w_2{\vi y}_1+w_1{\vi y}_2\mid \, \mid 
-w_1{\vi y}_1+w_2{\vi y}_2\mid \Big\}^{\ell+\ell'-\lambda},
\end{eqnarray}
and
\begin{eqnarray}
F_2({\vi y})&=&(x_1x_2)^{\lambda}
[Y_{\lambda}(\widehat{{\vi x}_1})Y_{\lambda}(\widehat{{\vi x}_2})]_{00}
\nonumber \\
&=&\frac{(-1)^{\lambda}\sqrt{2\lambda+1}}{4\pi}({\rm det}W)^{2\lambda}
\sum_{k=0}^{[\frac{\lambda}{2}]}(-1)^k
\frac{(2\lambda-2k-1)!!}{(\lambda-2k)!(2k)!!}
\nonumber \\
&\times& \Big\{(w_2{\vi y}_1+w_1{\vi y}_2)\cdot(-w_1{\vi y}_1+w_2{\vi y}_2)
\Big\}^{\lambda-2k}
\nonumber \\
&\times& \Big\{\mid w_2{\vi y}_1+w_1{\vi y}_2\mid \, \mid 
-w_1{\vi y}_1+w_2{\vi y}_2\mid \Big\}^{2k}.
\end{eqnarray}
Both $F_1({\vi y})$ and 
$F_2({\vi y})$ are polynomials of $y_1^2$,   $y_2^2$ and ${\vi y}_1\cdot 
{\vi y}_2$ as $\ell+\ell'-\lambda$ is an even integer, so that, with 
${\vi y}_2$ being replaced by ${\vi r}$, 
${\cal I}$ is reduced to the following type of elementary integrals 
\begin{equation}
\int {\rm e}^{-py_1^2+q{\vis r}\cdot{\vis y}_1}y_1^{2m}
({\vi r}\cdot {\vi y}_1)^{n}\, d{\vi y}_1,
\end{equation}
where both $m$ and $n$ are non-negative integers.

\end{document}